\documentclass{aastex}          
\usepackage{spr-astr-addons}    
\usepackage{graphicx}
\usepackage{etoolbox,lipsum}
\usepackage[colorlinks=true, linkcolor = blue,
urlcolor  = blue,
citecolor = blue,
anchorcolor = blue]{hyperref}

\makeatletter
\def\blfootnote{\xdef\@thefnmark{}\@footnotetext}
\makeatother

\RequirePackage{color}
\begin{document}
	%
	\title{Simulated differential observations of the Sunyaev-Zel'dovich Effect: Probing the Dark Ages and Epoch of Reionization}
	
	\shorttitle{<Short article title>}
	\shortauthors{<Autors et al.>}
	
	\author{C.M. Takalana\altaffilmark{1,2}} 
	\blfootnote{\href{mailto:mtakalana@ska.ac.za}{mtakalana@ska.ac.za}}
	\and 
	\author{P. Marchegiani\altaffilmark{1,3}}
	\blfootnote{\href{mailto:Paolo.Marchegiani@wits.ac.za}{Paolo.Marchegiani@wits.ac.za}}
	\and 
	\author{G. Beck\altaffilmark{1}}
	\blfootnote{\href{mailto:Geoffrey.Beck@wits.ac.za}{Geoffrey.Beck@wits.ac.za}}
	\and 
	\author{S. Colafrancesco\altaffilmark{1, $\dagger$}}
	\altaffiltext{$\dagger$}{Deceased: 30 September 2018}
	\altaffiltext{1}{School of Physics, University of the Witwatersrand, 1 Jan Smuts Avenue, Braamfontein, Johannesburg, 2000, South Africa}
	\altaffiltext{2}{South African Radio Astronomy Observatory, 2 Fir Street, Observatory 7925, South Africa}
	\altaffiltext{3}{Dipartimento di Fisica, Universit\`{a} La Sapienza, P. le A. Moro 2, Roma, Italy}


	\begin{abstract}
		This work presents an analytical approach for studying the cosmological 21cm background signal from the Dark Ages (DA) and subsequent Epoch of Reionization (EoR). We simulate differential observations of a galaxy cluster to demonstrate how these epochs can be studied with a specific form of the Sunyaev-Zel'dovich Effect called the SZE-21cm. This work produces simulated maps of the SZE-21cm and shows that the SZE-21cm can be extracted from future observations with low-frequency radio interferometers such as the Hydrogen Epoch of Reionization Array (HERA) and the Square Kilometre Array (SKA). In order to simulate near realistic scenarios, we look into cosmic variance noise, incorporate and take into account the effects of foregrounds, thermal noise, and angular resolution for our simulated observations. We further extend this exploration by averaging over a sample of galaxy clusters to mitigate the effects of cosmic variance and instrumental noise. The impact of point source contamination is also studied. Lastly, we apply this technique to the results of the EDGES collaboration, which in 2018 reported an absorption feature of the global 21cm background signal centred at 78 MHz. The challenges to be addressed in order to achieve the objectives of this work include errors that arise due to cosmic variation, instrumental noise and point source contamination.Our approach demonstrates the potential of the SZE-21cm as an indirect probe for the DA and EoR, and we conclude that the spectral features of the SZE-21cm from our simulated observations yield results that are close to prior theoretical predictions and that the SZE-21cm can be used to test the validity of the EDGES detection.
	\end{abstract}
	
	\section{Introduction} \label{sec:intro}
	The cosmic Dark Ages (DA) is an era of darkness in the Universe and the Cosmic Microwave Background (CMB) does not provide information about this period as baryonic matter and radiation have already decoupled prior to this epoch. Subsequently neutral hydrogen gas in the intergalactic medium (IGM) makes up the majority of baryonic matter in the Universe \citep{Zaroubi2013}. The DA end with the formation of the first generation of stars, that are believed to have had significant effects on the Universe and subsequently initiate the Epoch of Reionization (EoR) \citep{Bromm2009,Wise2019}. The first generation of stars and galaxies emit high-energy photons that permeate the IGM and gradually ionize the hydrogen, until at around $z \sim$  6 when the Universe is fully ionised \citep{Fan2006}. The process of reionization of the IGM enriches primordial galaxies with heavy elements, and seeds larger galaxies \citep{Furlanetto2006}. 
	\\
	Observations of the 21 cm hydrogen transition line promises to be an important probe into the DA and EoR. The 21 cm transition is a line, therefore the measured frequency corresponds to a particular redshift ($\nu$ = $\frac{\nu_{21}}{(1+z)}$ with $\nu_{21}$ = 1420.405 MHz), and by using this property we can isolate every phase in the history of the IGM. The signal is difficult to measure due to contamination by astronomical foregrounds \citep{Furlanetto2006}. Numerous efforts are nevertheless under-way to observe it and the experiments designed to measure it come in two different varieties namely power spectrum \citep{Paciga2013,Jacobs2015,Jacobs2016,DeBoer2017,Mondal2019} and global signal \citep{Burns2011,Bowman2009,Bowman2018}.
	\\
	The collaboration of the Experiment for the Detection of the Global EoR Signature (EDGES) recently reported the first observational results on the global 21 cm spectrum \citep{Bowman2018}. The results show an absorption feature with a remarkably large amplitude of about 500 mK at the redshift range of $z$ $\sim$ 15-20, which is nearly three times the predicted value. Until this detection is confirmed, it is essential that we evaluate the current theoretical models strictly and further explore the use of indirect techniques that can allow us to measure the 21 cm global background signal.
	\par
	\noindent
	Galaxy clusters are massive objects that provide powerful cosmological probes \citep{Romer2001}. Energetic electrons in the plasma of the galaxy clusters are responsible for inverse Compton scattering of background photons as they pass through the intracluster medium (ICM). This process produces a spectral distortion of the incident background radiation at the position of clusters on the sky and is known as the Sunyaev-Zel’dovich effect (SZE) \citep{SZ3,SZ1,SZ2,SZ}. The SZE has been identified as an effective cosmological probe and can be measured at microwave and radio frequencies \citep{Rephaeli1995,Birkinshaw1999,Holder1999}. 
	\\
	Existing interferometers cannot be used to detect the global average sky background temperature using standard techniques as they are not sensitive to the spatially invariant global average \citep{McKinley2018} but they are suitable to detect signal differences towards different directions in the sky with high precision, as in the case of the SZE. Based on this consideration, a specific form of the SZE called the SZE-21cm was proposed as an indirect technique to study the global 21 cm background signal \citep{Colafrancesco2016}. The SZE-21cm is a differential measurement towards and away from clusters of galaxies of the comptonized spectrum of the CMB as modified by physical processes occurring during the DA and the EoR. The idea to probe these epochs with the SZE-21cm was initially suggested by \cite{Cooray2006} and later addressed by  \cite{Colafrancesco2016}, who concluded that the effect can be observed at low radio frequencies (40 MHz-200 MHz) with upcoming interferometers such as the Murchison Widefield Array (MWA)\footnote{\url{http://www.mwatelescope.org}}, Low Frequency Array (LOFAR)\footnote{\url{http://lofar.org}}, Hydrogen Epoch of Reionization Array (HERA)\footnote{\url{http://reionization.org }}, and the Square Kilometre Array (SKA)\footnote{\url{http://www.skatelescope.org}}. 
	\par
	\noindent
	\cite{Colafrancesco2016} derived the SZE-21cm in a general relativistic approach and studied how its spectral shape can be used to establish the global features in the mean 21 cm spectrum generated during, and prior to, the EoR. Additionally, the aforementioned work also outlined how the effect depends on the properties of plasma in cosmic structures. \cite{Takalana2018} described the first version of a procedure to observe the SZE-21cm with upcoming interferometers. While the past work on this topic is mainly theoretical, in this work we perform new simulations of the background signal and study instrumental effects for HERA and SKA. We simulate differential observations of a low redshift galaxy cluster as a function of frequency for a theoretical 21 cm background model to show that by measuring the SZE-21cm we can derive the properties of the global 21 cm background spectrum generated during the DA and EoR. Our simulated observations explore the impact of cosmic variance (or sample variance) and point sources and take into account the effects of foregrounds, thermal noise, and angular resolution for upcoming interferometers. We also consider the observation and computation of the SZE-21cm towards multiple clusters as a method to ultimately reduce the uncertainty and improve the signal-to-noise. This work takes into consideration newer models of the 21 cm background, both theoretical and observational. 
	\\
	\noindent
	The EDGES experiment reported a 21 cm background signal that challenges cosmological models and instrumental precision. We use the EDGES detection results to make predictions for the SZE-21cm and make comparisons with the simulated theoretical 21 cm background model results to show that differential observation techniques for the SZE-21cm can be used to confirm detections made by experiments similar to EDGES. Lastly, we explore the feasibility of detecting the SZE-21cm with the upcoming SKA telescope.
	\par
	\noindent
	In section \ref{sec:theory} we present the theoretical background and mathematical formalism for studies of the 21 cm transition signal and the SZE-21cm. In section \ref{sec:standard} we simulate differential observations of a galaxy cluster with upcoming radio interferometers and from this establish the relationship between the SZE-21cm spectrum and the global 21 cm background spectrum using a theoretical 21 cm model. In section \ref{sec:Edges} we use the global 21 cm background signal measured by EDGES to make predictions for the SZE-21cm and make comparisons with the signal obtained through our simulations using the theoretical 21 cm model. In section \ref{sec:chal} we discuss the simulation results and look at possible challenges for observations of the SZE-21cm presented by instrumental effects, cosmic variance, and point sources. We summarize our findings and conclude in section \ref{sec:Conclusion}. We assume a  {$\Lambda$}CDM cosmology throughout this work with the following  Planck 2015 parameters: $h_{100}$ = 0.673, $\Omega_{m}$ = 0.315 ,  $\Omega_{b}$ = 0.0491, $\Omega_{\Lambda}$ = 0.685, $\sigma_{8}$ = 0.815 and $n_{s}$ =  0.968 \citep{Planck2015}.

	\section{The SZE-21cm: a probe for the DA and EoR}
	\label{sec:theory}
	
	In this section, we present theoretical background and formalism for studies of the 21 cm transition signal and discuss the detection of the signal claimed by the EDGES collaboration in 2018, and look at the concerns raised about the results. We also provide the theoretical background of the SZE-21cm and revisit the formalism derived by \cite{Colafrancesco2016}. Lastly, we discuss observations of the 21 cm to probe the DA and EoR.
	
	\begin{figure*}[t]
		\includegraphics[width=0.8\textwidth]{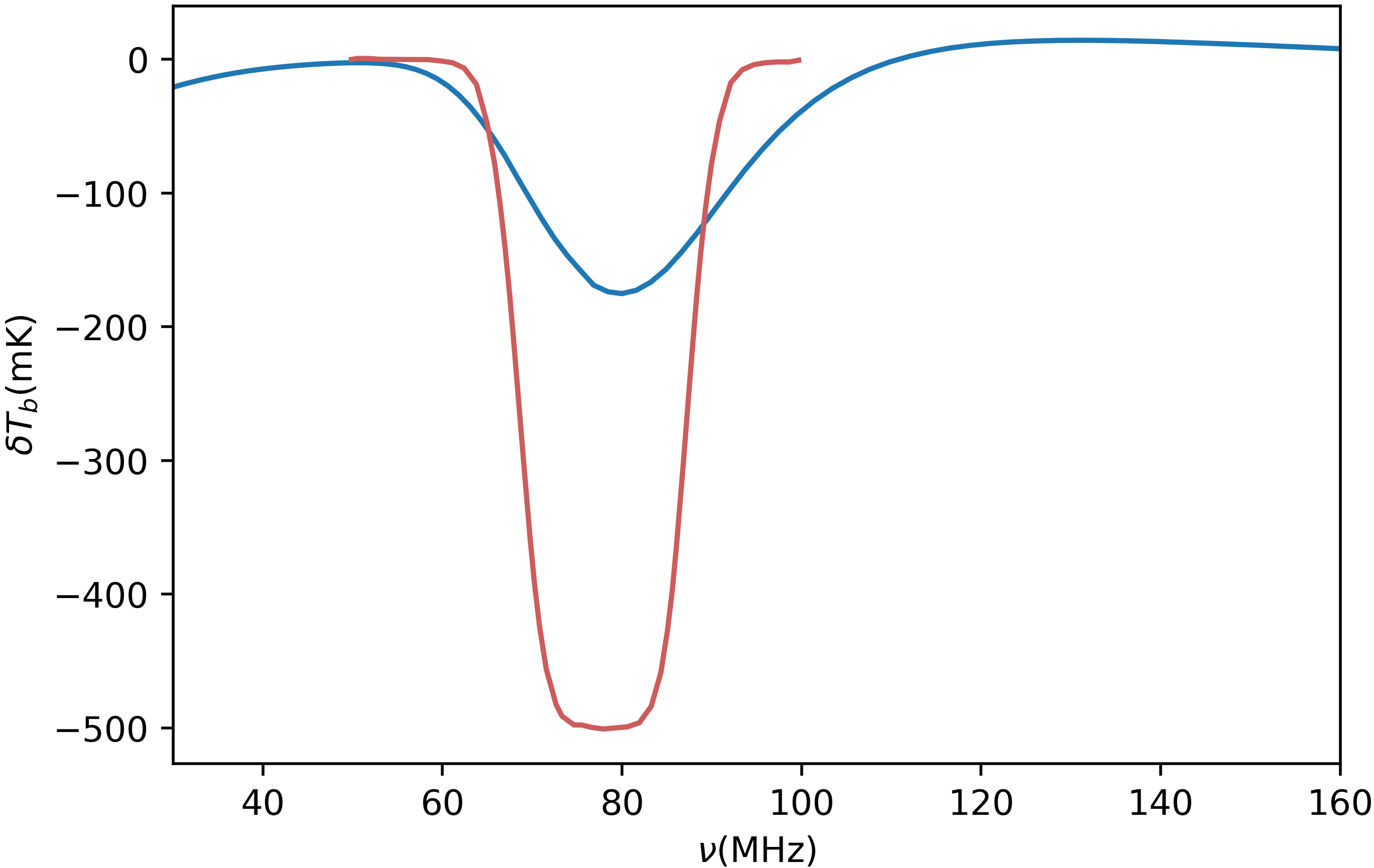}
		\centering
		\caption{The mean 21 cm background brightness temperature offset from the CMB spectrum as a function of frequency. The $\textbf{\emph{Blue}}$ curve is for a theoretical 21 cm model simulated for this work and the $\textbf{\emph{Red}}$ curve for the best-fitting 21 cm profile model derived from the EDGES observations \citep{Bowman2018}.}
		\label{fig:Tb}
	\end{figure*}
	
	\subsection{\textit{\textbf{The cosmological 21 cm background signal}}}
	\label{subsec:21cm}
	
	Neutral hydrogen atoms undergo a quantum mechanical transition between the hyperfine energy levels (spin-flip transition) resulting from magnetic interaction between the electron and the spin of the proton. When the spins change from parallel to anti-parallel, a photon with a wavelength of 21 cm is emitted, producing a signal with a line center frequency of $\nu_{0}$ = 1420.405 MHz \citep{vandeHulst1945, Ewen1951, Muller1957}. The intensity of the 21 cm radiation from a population of neutral hydrogen is dependent on the spin temperature ($T_{S}$) defined through the ratio between the number densities of hydrogen atoms in the $1S$ triplet and $1S$ singlet levels \citep{Pritchard2012}. The quantity of interest that radio interferometers will measure is the offset of the 21 cm brightness temperature from the CMB temperature, $T_{CMB} = 2.73(1+z)$ K, along a line of sight at the observed frequency \citep{Furlanetto2006}, which is written as

	\begin{multline}
		\hspace*{5mm}\delta T_b \simeq 23x_{HI} \left[\left(\frac{0.15}{\Omega_{m}}\right) \left(\frac{1+z}{10} \right) \right]^{1/2} \\
		\left(\frac{\Omega_{b}h^2}{0.02}\right) \left[1 - \frac{T_{R}}{T_{s}}\right] \textrm{mK},
		\label{eq:eq3}
	\end{multline}
	
	\noindent
	where $x_{HI}$ is the neutral fraction of hydrogen, $h$ is the Hubble parameter in units of 100 km s$^{-1}$ Mpc$^{-1}$,  and $\Omega_{b}$ and $\Omega_{m}$ are the baryon and matter densities respectively. $T_{R}$ is the temperature of the background radiation, which this work assumes to be from the CMB, hence $T_{R}$ = $T_{CMB}$. Predicted by \cite{Shaver1999}, this global 21 cm background signal is sensitive to both astrophysical and cosmological processes in the early Universe, which makes it a good probe of the physics between the CMB decoupling, the DA and the EoR \cite[see][for detailed discussions and formalism derivations for the 21 cm background signal]{Field1958,Shaver1999,Barkana2005,Furlanetto2006,Pritchard2012}. 
	
	\noindent
	This work focuses on a theoretical model of the 21 cm background signal that we simulate under standard cosmological assumptions using a semi-numeric modeling tool \cite{Mesinger2010} and one based on the best-fitting 21 cm profile model from recent EDGES results \citep{Bowman2018}. For both cases, the brightness temperature offset from the CMB is shown in Figure  \ref{fig:Tb} as a function of frequency. Equation \ref{eq:eq3} tells us that the amplitude of the absorption signal can be increased by reducing $T_{s}$. The claimed EDGES detection exhibits a much larger flat-bottomed amplitude of $-500_{-300}^{+200}$ mK with a Full-Width at Half-Maximum of $19_{-2}^{+4}$ MHz centred at around 78 MHz (corresponding to $z \sim 17$) \cite{Bowman2018}. Various proposals have been made to explain the larger than expected amplitude observed by EDGES. One of the proposals argues that the anomaly could be a new signal of baryon-dark matter interaction as scattering processes that may cool neutral hydrogen with respect to the standard expectations from the CMB measurements \cite[see][]{Barkana2018,Fialkov2018}. Another suggestion argues that the anomaly may be related to an additional radio background with likely causes of this excess ranging from instrumental systematics to new astrophysics as may have been observed by ARCADE 2 \cite[see][]{Fixsen2011,Ewall-Wice2018}. 
	\\
	\noindent
	Motivated by the unexpected nature of the EDGES claim, concerns were raised regarding the modelling of the EDGES data. These concerns arise from the following findings: 
	
	\begin{itemize}
		
		\item \cite{Hills2018,Singh2019,Sims22019} suggest the potential presence of an uncalibrated sinusoidal-form systematic in the data.
		
		\item \cite{Hills2018} also highlight the use of a modeling process that implies unphysical parameters in the data for foreground emission, where the degeneracy between signal and foreground modeling allows a wide range of signals to be consistent with the data. 
	
		\item Work by \cite{Singh2019} enforces a maximally smooth polynomial foreground model, constant amplitude sinusoidal systematic model, white noise and a Gaussian parametrisation of the global 21 cm signal, and found evidence for a different 21 cm signal, substantially more in agreement with the standard predictions.
		
		\item \cite{Bradley2019} report on a possible systematic artifact within the ground plane due to resonance modes beneath the antenna that may produce spurious broad absorption features in the spectra.

	\end{itemize}
	While the above ideas may not perfectly describe the detection, they illustrate the efforts of the community to use the results of \cite{Bowman2018} to explore possible exotic physics and new source populations as extensions to standard models for cosmology and particle physics. They also explore possible unmodeled systematic effects present in the data as we have seen above. Later in this work, we make predictions for the SZE-21cm using a 21 cm  background signal described by the \cite{Bowman2018} data to test the usefulness of the SZE-21cm as a technique that can potentially be used to verify the EDGES detection and other such claims in the future.
	
	\subsection{\textit{\textbf{The SZE-21cm}}}
	\label{subsec:sze21cm}
	
	The interaction between photons and free electrons by inverse Compton scattering results in the SZE \citep{SZ3, SZ1, SZ2, SZ}. The SZE is produced when the free electrons in the dense cores of galaxy clusters that contain hot ionized gas ($\sim$10$^7$ K) \citep{Cavaliere1976} up-scatter CMB photons causing a change in the apparent brightness of the CMB photons, which modifies the incident Planck spectrum \cite[see][for a review]{Birkinshaw1999}. This scattering results in a systematic shift of the CMB photons from the Rayleigh-Jeans to the Wien side of the spectrum \citep{Birkinshaw1999, Carlstrom2002, Colafrancesco2002}. 
	\\
	This work examines the use of low-frequency SZE observations ($<$ 200 MHz) to study the global properties of the universe during the DA and EoR with the 21 cm line of neutral hydrogen. In this section, we summarise the formalism for the SZE-21cm as derived by \cite{Colafrancesco2016} as this forms the basis of the simulated observations we present in the sections that follow. For the full treatment of the formalism in the relativistic limit, the reader is referred to \cite{Colafrancesco2002} for the standard SZE and \cite{Colafrancesco2016} for the SZE-21cm. Here we consider the SZE due only to the thermal electrons in a hot cluster, which requires the appropriate relativistic corrections to be taken into account. 	
	\par
	\noindent
	Any energy injecting process occurring during the EoR and DA modifies the CMB spectrum as described in Section \ref{subsec:21cm} and subsequently modifies the SZE. The incident CMB radiation field modified during the DA and EoR is given by the specific intensity as:
	
	\begin{align}
		\hspace*{5mm}I_{0,mod}(\nu) = I_{0,st}(\nu) + \delta I(\nu), 
		\label{eq:eq4}
	\end{align}
	
	\noindent
	where $\delta I (\nu)$  is the radiation intensity corresponding to the differential brightness temperature ($\delta T_b$) in equation \ref{eq:eq3} obtained by applying the Rayleigh-Jeans law, $\delta I (\nu) = \frac{2K_{B}\nu^{2}}{c^{2}} {\delta}T_{b}(\nu)$, and the unperturbed or standard CMB spectrum ($I_{0,st}$) is given by
	
	\begin{align}
		\hspace*{5mm}I_{0,st}(x) = 2 \frac{(k_{B}T_{CMB})^{3}}{(hc)^{2}} \frac{x^{3}}{e^{x}-1},
		\label{eq:eq5}
	\end{align}
	
	\noindent
	where $c$ is the speed of light, $h$ is the Planck constant, $k_{B}$ is the Boltzmann constant, and ${x = \frac{h\nu}{k_{B}T_{CMB}}}$ is the a-dimensional frequency. The spectral distortion due to the SZE of the CMB radiation field as modified during the DA and EoR is given by
	
	\begin{align}
		\hspace*{5mm}I_{mod}(x) = \int_{-\infty}^{+\infty} I_{0,mod}(xe^{-s}) P(s) ds,
		\label{eq:eq6}
	\end{align}
	
	\noindent
	where $P(s)$ is the photon redistribution function that is dependent on the electrons spectrum producing the CMB Comptonization \citep{Birkinshaw1999,Ensslin2000,Colafrancesco2002,Colafrancesco2016}, and is essentially the scattering kernel giving the probability of a scattering to shift a photon from a frequency $\nu$ to $\nu'$ (with $s= \ln(\frac{\nu'}{\nu})$). The SZE-21cm is then given as the difference between the scattered spectrum ($I_{mod}$) and the incoming radiation ($I_{0,mod}$):
	
	\begin{align}
		\hspace*{5mm}\Delta I_{mod}(\nu) = I_{mod}(\nu) - I_{0,mod}(\nu).
		\label{eq:eq7}
	\end{align}
	
	\noindent
	This equation can then be transformed in terms of brightness temperature by applying the Rayleigh-Jeans law and be rewritten as
	
	\begin{align}
		\hspace*{5mm}\Delta T_{mod}(\nu) = T_{mod}(\nu) - T_{0,mod}(\nu).
		\label{eq:eq8}
	\end{align}
	
	\subsection{\textit{\textbf{Observations of the 21 cm signal from the DA and EoR}}}
	\label{subsec:observations}
	
	The 21 cm background signal from the DA and EoR is weak and may be observed as a faint, diffuse background detectable at frequencies below 200 MHz. Previous and current generation interferometer experiments attempt to measure a statistical power spectrum of the signal over the sky, rather than to image the signal directly, by measuring a range of aggregate spatial scales on the sky. There are several challenges associated with observations of the 21 cm background. The main problem is that the signal is weak, at a characteristic scale of $\sim$ 10 to 100 mK, and the task is further complicated by the much stronger foreground continuum emission due to sources such as galactic diffuse synchrotron, free-free emission, and emissions from extragalactic radio sources \citep{Santos2005,Sims2019}. These contaminating signals are five orders of magnitude stronger than the 21 cm emission, making its observation extremely complex to study, and requiring very precise calibration and knowledge of various foregrounds \cite[see e.g.,][]{Furlanetto2006}. Several ideas have been presented to overcome the various problems identified above, including foreground avoidance and removal techniques. These strategies rely on the spectral smoothness and spatial characteristics of foreground sources; we do not discuss the various approaches here but refer the reader to other texts \cite[see e.g.,][]{Barkana2005,Furlanetto2006,McQuinn2006}. 
	\\
	\noindent
	All observations that are sufficiently sensitive to detect the SZE are differential \citep{Carlstrom2002}. Consequently, differential observations of the SZE-21cm towards galaxy clusters have been proposed as one way to overcome the above-mentioned problems by \cite{Cooray2006} and \cite{Colafrancesco2016}. Unlike an experiment to directly determine the cosmological 21 cm background spectrum involving a total intensity measurement on the sky, differential observations with an interferometer are less affected by the confusion from galactic foregrounds that are smooth over angular scales larger than a typical cluster and by issues such as the exact calibration of the observed intensity using an external source \citep{Cooray2006,Colafrancesco2016}. In the next section, we simulate differential observations of a galaxy cluster to compute the SZE-21cm.
	
	\section{Simulations of differential observations of the SZE-21cm}
	\label{sec:standard}
	\subsection{\textit{\textbf{Simulations of the 21 cm background}}}
	We simulate the 21 cm cosmological background using the publicly available semi-numerical code 21cmFAST\footnote{\url{https://github.com/andreimesinger/21cmFAST}}. 21cmFAST takes into account the necessary physical processes and produces large-scale simulations utilizing perturbation theory, the excursion set formalism (which specifies the location and mass of collapsed dark matter haloes), and analytic prescriptions to produce evolved 3D realizations of density, spin temperature fields, ionization, and peculiar velocity, which when combined, can be used to deduce the global 21 cm brightness temperature at a given redshift as described by equation \ref{eq:eq3}. 
	\\
	Each simulation cube in our runs is 30 Mpc on a side down-sampled from the high-resolution initial conditions generated on a 460$^{3}$ grid and computed on a 115$^{3}$ grid corresponding to a cell resolution of 0.26 Mpc, in the frequency range $\nu$ = 30 - 160 MHz. Our model assumes that Population II stars are responsible for early heating with parameters in Table \ref{table:tb1}. The parameters are the ionizing photons per stellar baryon $N_{\gamma}$, the ionizing efficiency (the number of photons per baryon per dark matter halo) $\zeta$, the fraction of baryons converted to stars $f_{*}$, the minimum virial temperature for all sources $T_{vir,min}$, the maximum horizon for ionizing photons $R_{max}$, the efficiency parameter corresponding to the number of photons per solar mass in stars ${\zeta}_{X}$, and the clumping factor on the scale of the simulation cell $C$. We refer the reader to \cite{Mesinger2010} for detailed information and discussions on the code, relevant parameters, and the formalism.
	
	\begin{table}[htbp]
	\caption{Summary of parameters used in our run of the 21cmFAST simulation run.
	\label{table:tb1}}
	\centering
	\begin{tabular}{ c c c }
	\tableline\tableline
	Parameter & \\ 
	\tableline
	$N_{\gamma}$ & 4300 \\
	$\zeta$ & 20 \\ 
	$f_{*}$ & 0.05 \\
	$T_{vir,min}$ & $10^{4}$ K \\
	$R_{max}$ & 50 Mpc \\
	${\zeta}_{X}$ & $10^{56} \ M_{\odot}$\\
	$C$ & 2 \\
	\tableline\tableline
	\end{tabular}
	\end{table}
	
	\subsection{\textit{\textbf{Procedure for simulated observations of the SZE-21cm}}}
	\begin{figure}[t]
		\includegraphics[width=0.483832\textwidth]{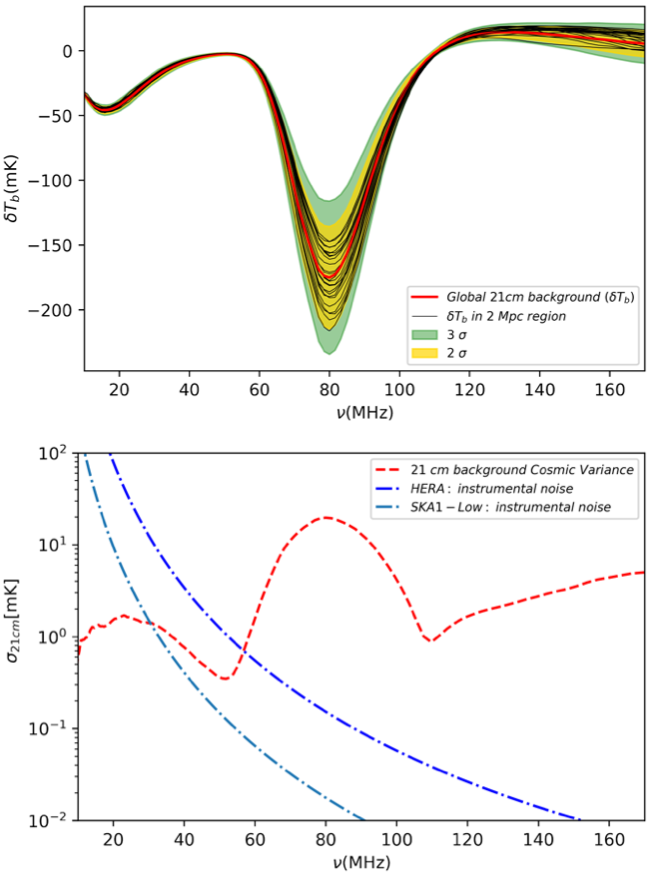}
		\centering
		\caption{\textbf{Top}: The mean 21 cm background brightness temperature offset from the CMB spectrum as a function of frequency incident on each of the cluster-sized regions ($\textbf{\emph{Black}}$), along with the simulation global mean ($\textbf{\emph{Red}}$). The shaded areas correspond to 2$\sigma$ and 3$\sigma$ standard deviations in the cluster-sized regions. \textbf{Bottom}: Noise on the 21 cm Global signal as a function of frequency (y-axis in log-scale). The ($\textbf{\emph{Red}}$)-dashed lines represent the cosmic-variance.  For comparison, we also have the instrumental noise for the HERA and SKA1-Low experiments calculated using equation \ref{eq:eq15}.}
		\label{fig:var2}
	\end{figure}
	This work makes use of only the 21 cm brightness temperature ${\delta}T_{b}(\nu)$ cubes from the 21cmFAST simulations. We use equation \ref{eq:eq5} to compute unperturbed CMB intensity cubes ($I_{0,st}(\nu)$), and Rayleigh-Jeans law to calculate it in terms of brightness temperature ($T_{0,st}(\nu)$), and add this signal to each cube of ${\delta}T_{b}(\nu)$, which gives us the CMB radiation field modified during the DA and EoR ($T_{0,mod}(\nu)$). To conduct our mock observations of the SZE-21cm we insert a typical rich cluster with a temperature of 10 keV, an optical depth of 10$^{-3}$, and a radius of 2 Mpc. The incoming background radiation field  ($T_{0,mod}(\nu)$) is scattered by energetic electrons in the cluster, resulting in a modified signal ($T_{mod}(\nu)$) calculated according to equations \ref{eq:eq6} - \ref{eq:eq8}. 
	\\
	\noindent
	One of the issues that we will face measuring the SZE-21cm is that the 21 cm background has variations on angular scales comparable to that of the galaxy clusters and their intrinsic level presents a potentially significant contaminant to SZE-21cm observations as the variations may be comparable to the expected signal of the SZE-21cm as computed by \cite{Colafrancesco2016}. The variations ($\frac{\Delta T}{T}$) in our cubes are computed to be of orders between 10$^{-1}$ - 10$^{-3}$. They arise from significant degeneracies between parameters that control the average 21 cm global temperature spectrum, which lead to inhomogeneities \citep{McQuinn2006}, and are greater than the CMB primordial fluctuations that are of the order of 10$^{-5}$. To quantify the cosmic variance noise that arises from the finite volume of the universe accessible to 21 cm experiments we plot the 21 cm signal (versus frequency) incident on each of the cluster-sized angular regions in the simulations, along with the global average (see Figure \ref{fig:var2}). The figure illustrates how the global signal measured over the cluster-sized region can significantly depart from the average signal. Each of the cluster-sized regions provides an estimator for the ${\delta}T_{b}(\nu)$, which varies from one to another, showing variance produced to the signal in a finite region. We note that cosmic variance noise increases the error budget resulting in a decrease in the significance of any detection. Hence, by allowing observations of many more samples we will reduce the overall error budget and effects of cosmic variance.
	\\
	\noindent
	Observations of the SZE at microwave frequencies will aid in distinguishing the SZE-21cm fluctuations from intrinsic degeneracies between parameters that control the mean 21 cm brightness temperature spectrum and CMB fluctuations, and we can use information from these observations as a way to normalise low-frequency signals that will be crucial to establishing signatures related to the 21cm signal. Contamination from point sources is also a major concern and for our mock observations, it is, therefore, useful that we detect and discard point-source contaminated pixels. To do this we apply a threshold approach to avoid contamination in the SZE-21cm computation, identifying possible point sources in these fields above a given threshold and replacing those pixels with the mean background value; this also helps us account for the cube fluctuations. Later in the section \ref{sec:chal}, we will focus on how possible sources of contamination can be handled. 
	
		\begin{figure*}
		\includegraphics[width=0.99\textwidth]{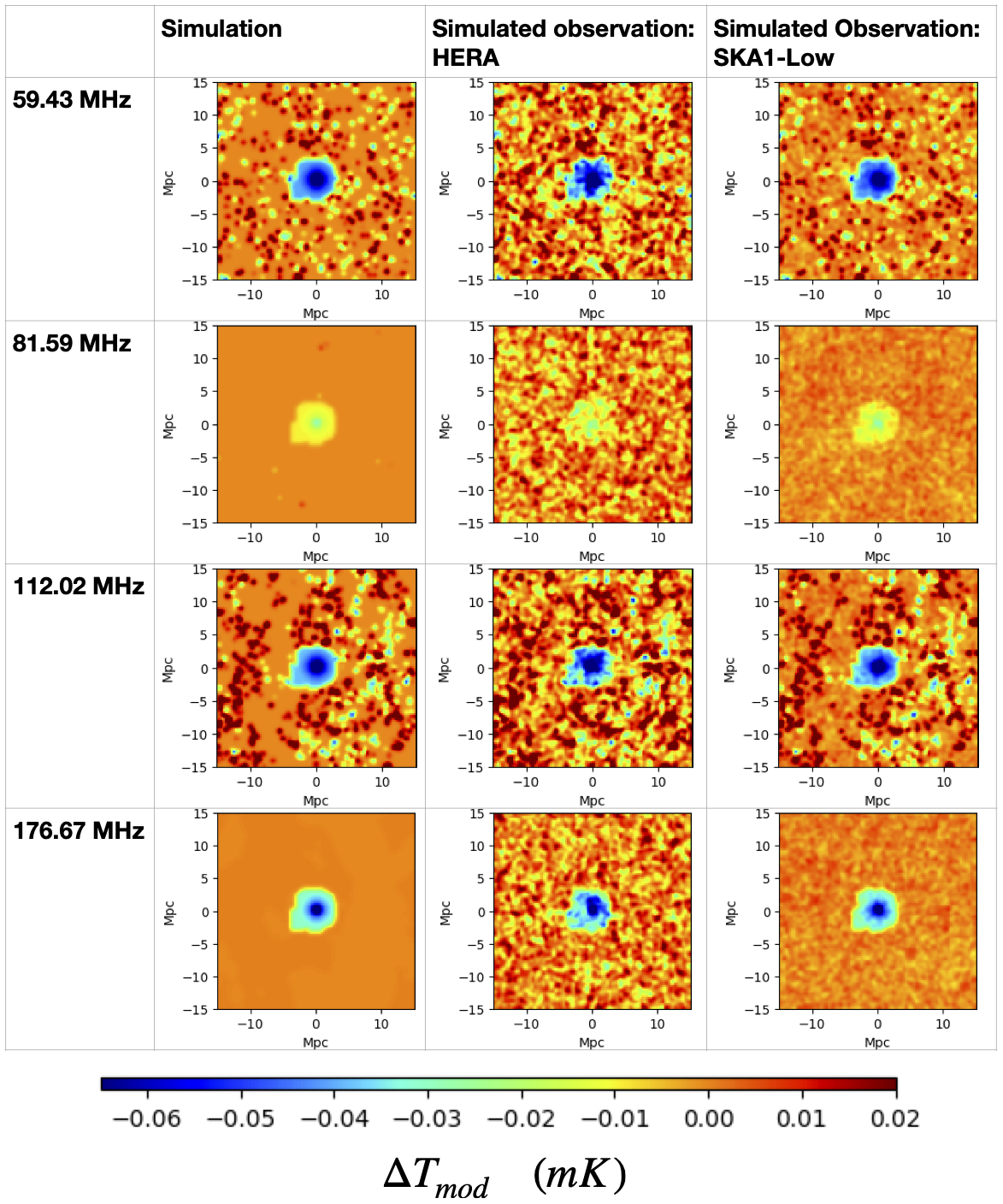}
		\centering
		\caption{Simulated SZE-21cm cluster maps from our mock differential observations at 59.43 MHz, 81.59 MHz, 112.02 MHz, and 176.67 MHz for a cluster with a plasma temperature of 10 keV, a radius of 2 Mpc, and optical depth of 10$^{-3}$. The maps are shown for simulated differential observations without instrumental effects and simulated observations with HERA and SKA1-Low.}
		\label{fig:image}
	\end{figure*}
	
	\begin{table*}[t]
	\caption{Summary of parameters used in this work for HERA and SKA.}
	\label{table:tb2}
	\centering
	\begin{tabular}{ c c c }
	\tableline\tableline
	 & \textbf{HERA} & \textbf{SKA1-low} \\ 
	\tableline
	Antennae design & 350 hexagonally packed dishes & 512 stations \\
	Antenna/station diameter (m) & 14 & 35 \\ 
	Collecting area ($m^{2}$) & 53,878 & 416,595 \\
	$A_{eff}$ ($m^{2}$) & 153.93 & 962.11 \\
	$T_{sys}(K) (= T_{sky} + T_{rcvr})$ & $T_{sky} + 100$ & $1.1T_{sky} + 40$ \\
	\tableline\tableline
	\end{tabular}
	\end{table*}
	
	\subsubsection{Simulating 21cm instruments}
	
	To include realistic 21 cm instrumental effects we follow the methodology of \cite{Hassan2018} that accounts for the finite angular resolution of the instrument, foreground cleaning, and the presence of instrumental noise. This methodology produces mock observations using the 21cmSense\footnote{\url{https://github.com/jpober/21cmSense}} package \citep{Pober2013,Pober2014}, which creates realisations of thermal noise taking into account the antenna configuration, system temperature, and station/dish size to calculate expected sensitivities of 21 cm experiments. For this work, we produce mock observations with HERA and the low-frequency SKA1 (SKA1-low) instruments. The summary of our assumed HERA and SKA array designs are in Table \ref{table:tb2}.
	
		\begin{figure}[t]
		\includegraphics[width=0.4838\textwidth]{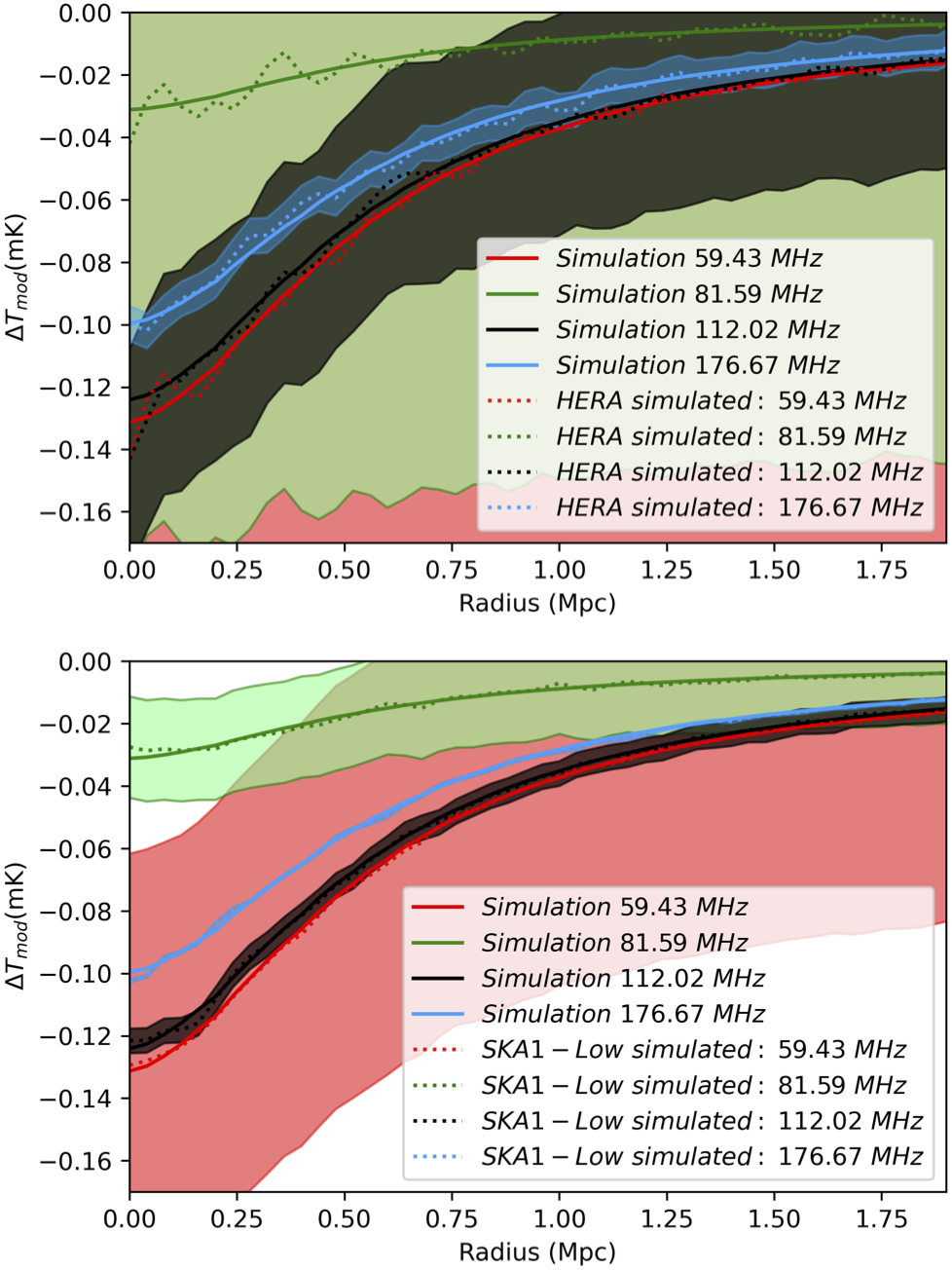}
		\centering
		\caption{The SZE-21cm temperature decrement radial profile from a cluster of radius of 2 Mpc, a core radius ($r_c$) of 0.248 Mpc and $\beta$ = $\frac{3}{5}$ at 59.43 MHz, 81.59 MHz, 112.02 MHz, and 176.67 MHz. \textbf{Top}: simulation without instrumental effects and simulated observations with HERA. \textbf{Bottom}: simulation without instrumental effects and simulated observations with SKA1-Low. The shaded errors regions assume observations with HERA \textbf{(Top)} and SKA1-Low \textbf{(Bottom)}.}
		\label{fig:rad}
	\end{figure}

	 \noindent
    We take a 3D-Fourier transform of each cube, this gives us the magnitude of the cube as a function of wave vector \textbf{k}; $k$, the magnitude of \textbf{k} vector, is $k = |\textbf{k}| = \sqrt{k_{\perp}^{2}+k_{\parallel}^{2}}$, where $k_{\perp}$ is the co-moving component on the plane of the sky and  $k_{\parallel}$ is the co-moving component along the line of sight. It is expected that foreground contamination of 21 cm measurements will be spectrally smooth and should only affect the lowest $k_{\parallel}$ modes \citep{Bowman2009}. However, due to the frequency dependence of an interferometer’s response, the contamination will leak into higher $k$ modes through ``mode mixing", which is stronger for long baselines (high values of $k_{\perp}$) that have higher fringe rates. This mode mixing causes the foregrounds to become confined to a wedge-shaped region in the $k_{\perp} - k_{\parallel}$ space \citep{Liu2009,Liu2014}.
    \\
    \noindent
    First, we remove the foreground contaminated $k$ modes from the cubes assuming that foregrounds are confined to some regions in Fourier space along the line-of-sight. This confinement is expressed in terms of spatial Fourier wave-numbers for Fourier modes $k_{\perp}$ and $k_{\parallel}$. Foreground contaminants are expected to be mostly in modes that satisfy the following condition:
    
    \begin{align}
    \hspace*{5mm}k_{\parallel} < k_{\parallel}^{0} \frac{H_{0}\:D_{c}\: \theta[\Omega_{m} \:(1+z)^{3} + \Omega_{\Lambda}]^{\frac{1}{2}}}{c(1+z)} k_{\perp},
    \label{eq:eq9}
    \end{align}
    \noindent
    where  $D_{c}$ is the co-moving line-of-sight distance, $\theta$ is the angle between the line of sight and the direction of the Fourier mode \textbf{k}, and $k_{\parallel}^{0}$ is an offset constant \cite[detailed derivations of this may be found in][]{Liu2014,Liu2016}. The modes of the signal that are contaminated by foregrounds lie inside a foreground wedge in the $k_{\perp} - k_{\parallel}$ plane. The foreground wedge slope is given by:
    
    \begin{align}
    \hspace*{5mm}m = \frac{H_{0}\: D_{c}\: [\Omega_{m} \:(1+z)^{3} + \Omega_{\Lambda}]^{\frac{1}{2}} \:\textrm{sin}\theta}{c(1+z)}.
    \label{eq:eq10}
    \end{align}

    \noindent
    For our cubes, we zero out all the modes that are inside the wedge to clean the foregrounds. More modes are removed at lower frequencies since $m$ increases with decreasing frequency. To account for the angular resolution we compute the uv-coverage antennae distribution for the HERA and SKA1-Low designs using the 21cmSense package. The uv-coverage gives the total number of baselines that observe a given u-v pixel. The angular direction is given by $k_{x} = 2\pi k_{y}$, where the sky along x and y coordinates can be converted in terms of baseline length in u and v coordinates. We compute the u-v coverage for HERA and SKA1-Low from the antenna distribution and convert the uv coordinates into corresponding $k_{x}$ and $k_{y}$ modes. We then obtain a Fourier transform of $T_{mod}(\nu)$ and $T_{0,mod}(\nu)$ maps, and zero out the Fourier modes that correspond to null u-v coverage at $k_{x}$ and $k_{y}$. We smooth down the maps using a Gaussian filter with a full width half maximum of FWHM = $\frac{21cm \: (1+z)}{B}$, where $B$ is the maximum baseline length for HERA and SKA1-Low. Thereafter, we obtain the angular resolution limited $T_{mod}(\nu)$ and $T_{0,mod}(\nu)$ maps and we inverse Fourier transform back to real space. We simulate a noise map whose pixel values are obtained from a Gaussian distribution with a zero mean and standard deviation that is set to thermal noise for HERA and SKA1-Low \citep[see,][]{Zaldarriaga2004,Hassan2018}:
    
    \begin{align}
    \hspace*{5mm}\sqrt{\langle |N|^{2} \rangle} [Jy]= \frac{2k_{B}T_{sys}}{A_{eff}\sqrt{\Delta \nu \: t_{int}}},
    \label{eq:eq11}
    \end{align}
    \noindent
    where $A_{eff}$ is the effective area of a single antenna, $\Delta \nu$ is the frequency resolution, $t_{int}$ is the integration time, and $T_{sys} = T_{sky} + T_{rcvr}$ is the system temperature, where $T_{rcvr}$ is the receiver temperature and the sky temperature is, $T_{sky} = 120 \big(\frac{\nu}{150}\big)^{-2.55} \textrm{K}$, see Table \ref{table:tb2}. Our simulated experiments assume $t_{int}$ = 1000 hours, and  $\Delta \nu$ = 97.8 kHz for HERA and $\Delta \nu$ = 50 kHz for SKA. The noise is suppressed by our u-v coverage $(N_{uv})$ by a factor of $\frac{1}{\sqrt{N_{uv}}}$, we inverse Fourier transform the noise map to real space. This is then added to our foreground and angular resolution corrected $T_{mod}(\nu)$ and $T_{0,mod}(\nu)$ maps. 
    
    \subsubsection{Mock differential observation}
	We perform mock observations in the direction of the cluster $T_{mod}(\nu)$ and compare this with an empty sky region located away from the cluster region $T_{0,mod}(\nu)$ to obtain SZE-21cm maps with HERA and SKA1-Low. The empty sky region away from the cluster has the same size as the region of the cluster to allow for pixel-by-pixel operations. The differential observation that gives us the SZE-21cm as in equation \ref{eq:eq8} can be written as: 
	
	\begin{align}
		\hspace*{5mm}\Delta T_{SZE-21cm} = T_{ON} - T_{OFF},
		\label{eq:eq12}
	\end{align}
	
	\noindent
	where $T_{ON}$ is the value obtained when observing the cluster signal (hereafter titled signal ON), from which we subtract the background (empty sky), $T_{OFF}$ (hereafter titled signal OFF). To account for correlation and position-dependent offsets, we use two background regions and let $T_{OFF} = \frac{T_{OFF_{1}} +T_{OFF_{2}}}{2}$. We perform the differential observation as set out by equation \ref{eq:eq12}, which gives us the SZE-21cm at the cluster as $\Delta T_{mod} = \Delta T_{SZE-21cm}$, which is the modification of the initial background signal. Assuming that the mock cluster's gas is isothermal, we use the isothermal $\beta$-model to model its density profile \cite[see e.g.,][]{Cavaliere1976}, which varies as a function of cluster radius, $r$:
	
	\begin{align}
		\hspace*{5mm}n_{e}(r) = n_{e,0} \bigg(1 + \frac{r^{2}}{r_{c}^{2}}\bigg)^{-\frac{3}{2} \beta},
		\label{eq:eq13}
	\end{align}
	
	\noindent
	where $n_{e,0}$ is the electron number density at the center of the spherical symmetric cluster gas distribution; $\beta$ is the power-law index and $r_{c}$ is the core radius. This electron density distribution leads to the following analytic form of SZE-21cm brightness temperature profile: 
	
	\begin{align}
		\hspace*{5mm}\Delta T_{mod}(\nu) = \Delta T_{mod,0} \bigg(1 + \frac{r^{2}}{r_{c}^{2}}\bigg)^{\frac{1}{2} - \frac{3}{2} \beta},
		\label{eq:eq14}
	\end{align}

	\begin{figure*}[t]
	\includegraphics[width=0.7\textwidth]{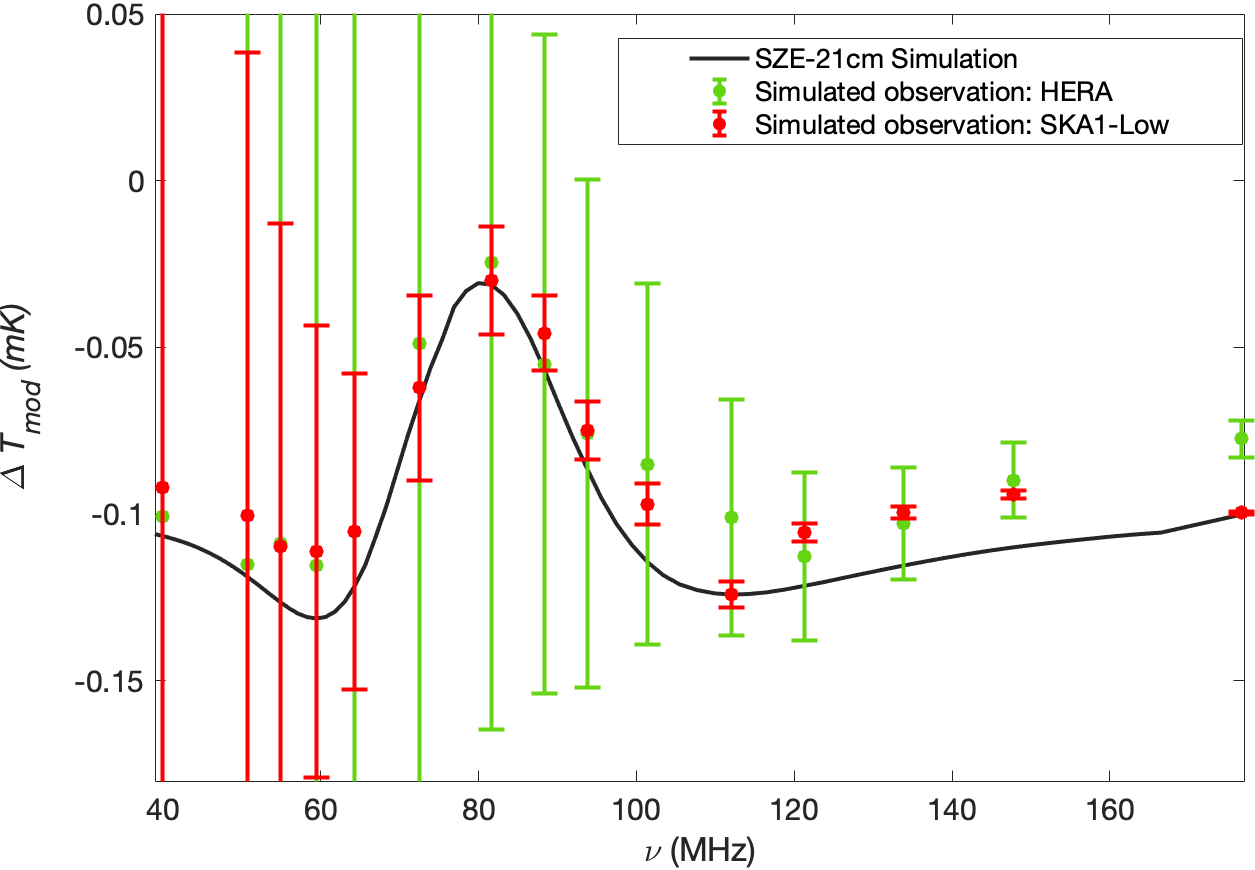}
	\centering
	\caption{The SZE-21cm spectrum from our mock differential observations of a galaxy cluster with a plasma temperature 10 keV and {$\tau = 10^{-3}$}. The \textbf{Black} solid line is the theoretically expected value calculated for our simulated cluster. The \textbf{Green} points and error bars are from the SZE-21cm measurement of the simulated observations for HERA, and the \textbf{Red} points and error bars are from the SZE-21cm measurement of the simulated observations for SKA1-Low.}
	\label{fig:HERASKASZE}
	\end{figure*}
	
	\noindent
	$\Delta T_{mod,0}$ is the central temperature decrement. We model the SZE-21cm brightness temperature, $\Delta T_{mod}$, at the cluster using the model in equation \ref{eq:eq14}, with $\beta$ = $\frac{3}{5}$ and $r_{c}$ = 0.248 Mpc for a typical rich cluster. 
	\\
	We generate cluster maps and measure the SZE-21cm at the cluster, and use the measured signal to construct the SZE-21cm spectrum. The simulated maps are shown in Figure \ref{fig:image} and the cluster radial profiles are in Figure \ref{fig:rad}. The mock observations are made in the frequency range between 30 and 160 MHz, but here we only show the simulated maps and cluster radial profiles at at 59.43 MHz, 81.59 MHz, 112.02 MHz, and 176.67 MHz as these are the frequencies where we can see the strongest decrements or increments in the SZE-21cm signal. The SZE-21cm spectrum is shown in Figure \ref{fig:HERASKASZE} and is discussed at length in section \ref{disc}.

	\subsection{\textit{\textbf{Discussion: HERA and SKA simulated observations of the SZE-21cm for the theoretical model}}}
	\label{disc}

	According to \cite{Cooray2006} and \cite{Colafrancesco2016}, detailed measurement of the SZE-21cm will allow us to derive precise information on the epochs at which the CMB was modified (i.e., DA and EoR) and on the physical mechanisms that imprint distinct features on the global 21 cm background ($\delta T_{b}$) that are present in the SZE-21cm spectrum. Figure \ref{fig:image} in the work presented here shows images resulting from the simulated differential observations of a galaxy cluster at different frequencies, and figure \ref{fig:HERASKASZE} gives us the SZE-21cm measured at the cluster without instrumental effects and also taking into account the effects of foregrounds, thermal noise, and angular resolution observing with HERA and SKA1-Low, which gives us a good idea of the impact that instrumental effects have on the differential observations. For our simulated observations using a low-frequency radio interferometer, we employ a simple strategy to provide order of magnitude approximations to the noise errors at the frequencies of relevance. The system temperature ($T_{sys}$) for HERA and SKA1-Low can be found in Table \ref{table:tb2}. The instrumental noise on the brightness temperature, $\Delta T^{N}$, measured by an interferometer is given by \cite{Furlanetto2006}
	
	\begin{align}
		\hspace*{5mm}\Delta T^{N} = \frac{T_{sys}}{\eta_{f} \sqrt{\Delta \nu t_{int}}}.
		\label{eq:eq15}
	\end{align}
	
	\noindent
	The array filling factor is given by $\eta_{f} = {A_{eff}}/D^{2}_{max}$, where $D_{max}$ and $A_{eff}$ are the maximum baseline and total effective area of the array respectively.
	\\
	\noindent
	We realise that the SZE-21cm spectrum almost resembles an inverted $\delta T_{b}$ spectrum. As the SZE-21cm changes as a function of frequency, its features correspond to some changes in the $\delta T_{b}$ spectrum. For example, the minimum point of the input spectrum at approximately 80 MHz, Figure \ref{fig:Tb}, corresponds to the maximum point of the SZE-21cm spectrum, Figure \ref{fig:HERASKASZE}. The spectrum shows absorption of the SZE-21cm between 40-60 MHz, as a result, we observe a decrement in the SZE-21cm map at 59.43 MHz, this followed by emission in the SZE-21cm spectrum between 60-90 MHz and as a result, we observe an increment on the SZE-21cm map at 81.59 MHz. Between 80-110 MHz there is absorption in the SZE-21cm spectrum and as a result, we observe another decrement at 112.02 MHz. At $\nu >$ 110 MHz the SZE-21cm spectrum starts to pick-up and this is evident in the SZE-21cm map at 176.67 MHz. Notably, we have a deviation between the simulated observation and the true signal at high frequencies in Figure \ref{fig:HERASKASZE}, and the deviation is more substantial for SKA1-Low than for HERA, contrary to expectations, and it is not clear what produced this effect.
	\\
	The cluster is easily identifiable in all the simulated maps thanks to the angular resolution of our simulated HERA and SKA1-Low designs and the uv-coverage that extends small scales at the frequencies of interest. The noise for SKA1-Low is less as compared to HERA and the lower angular resolution for HERA makes it slightly more challenging to resolve the cluster. Figure \ref{fig:rad} also shows that we can recover the cluster radial profile with both HERA and SKA as the simulated observation results with instrumental effects are compatible with the model. At higher frequencies, SKA1-Low will be able to produce good results, however, at low frequencies, the errors are larger. SKA1-Low may be an ideal instrument to conduct studies of the SZE-21cm as compared to HERA, which promises to be instrumental in making the first detections of the effect as it will come online ahead of SKA1-Low. Minimising the errors when observing with HERA will require significant amount of integration time, this work simulate the observations assuming 1000 hours with both instruments.
	
	\subsubsection{\textit{The SZE-21cm from multiple clusters}}
\begin{figure*}[t]
		\includegraphics[width=0.7\textwidth]{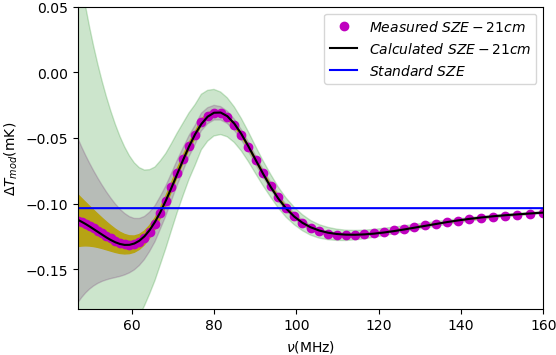}
		\centering
		\caption{The SZE-21cm spectrum from our mock differential observations of a galaxy cluster with a plasma temperature 10 keV and {$\tau = 10^{-3}$}. The shaded errors regions assume observations with an interferometer similar to SKA. The $\textbf{\emph{Green}}$ error region is for differential observations towards a single cluster, $\textbf{\emph{Grey}}$ area towards 10, and the $\textbf{\emph{Gold}}$ area towards 100 clusters \cite[see e.g.,][]{Cooray2006}. The $\textbf{\emph{Black}}$ solid line is the best-fit to theoretically calculated data points, the $\textbf{\emph{Magenta}}$ dots are points measured from our simulated mock observations and the $\textbf{\emph{Blue}}$ solid line is the standard SZE ($\Delta T_{st}$) for comparison.}
		\label{fig:sze21spec}
	\end{figure*}

	The SZE-21cm, which is our quantity of interest, is a faint signal and the best way to deal with such a signal would be to average it over as many clusters as possible. We demonstrate the feasibility of reducing the error budget that arises due to instrumental effects by averaging the SZE-21cm observed towards multiple galaxy clusters with the low-frequency SKA telescope, we estimate the expected error for cluster observations of the SZE-21cm by scaling and combining observations towards multiple clusters to improve the signal-to-noise which is given by equation \ref{eq:eq15}. The expected errors in Figure \ref{fig:sze21spec} are determined assuming observations towards one cluster and the improvement by computing the average over signals towards ten and then one hundred clusters. To give a sense of the impact this may have on reducing the error budget, we assume an improvement in $\Delta$ $T$  with $\sqrt{N_{cl}}$ of the number of clusters $(N_{cl})$ used \citep{Cooray2006,Clanton2012}. In each of the clusters, the SZE-21cm signal is present, thereby adding to it its contribution. Whereas, noise is random and therefore does not add but begins to cancel as the number of clusters we average increases. For this reason, we can expect that the signal-to-noise improvement obtained by averaging the SZE-21cm from multiple galaxy clusters to be proportional to the square root of the number of clusters averaged. Measuring the  SZE-21cm observed towards multiple galaxy clusters and obtaining an average as described here will also be beneficial to reduce the uncertainty that arises due to cosmic variance on the 21 cm background signal incident on each of the cluster-sized angular regions in the simulations, see Figure \ref{fig:var2}, which we begin to suppress stacking up to a few hundred clusters.
	
	 \noindent
	The SZE-21cm computation requires prior knowledge of the cluster's electron temperature and optical depth profile, which can be obtained from observations of the SZE related to the CMB spectrum alone at much higher radio and microwave frequencies, possible with the Planck telescope \cite[see e.g.,][]{Hansen2002}. \cite{Colafrancesco2016} pointed out that the SZE-21cm as a function of frequency is dependent on the temperature of the intracluster plasma, mathematically this can be written in terms of the Compton parameter as $y \propto k_{B}T \cdot \tau$. We propose two solutions to overcome the challenge that arises from the SZE-21cm varying for clusters with different plasma temperatures. To average the SZE-21cm of clusters with different plasma temperatures, we can study and measure the Compton $y$ parameter at higher frequencies and factor it in our computations, alternatively we can average the SZE-21cm for clusters with the same plasma temperatures only. In this regard, we expect the low-frequency SKA array to have large fields-of-view of approximately 2.5 – 10 degree from 200 MHz down to 50 MHz \citep{Mellema2012}, which will enable us to identify large cluster samples to study the SZE-21cm by selecting the ones having similar plasma temperatures as can be derived from X-ray observations. Detecting a larger number of sources will also strengthen the statistical studies of the DA and EoR and the cosmological radio backgrounds. 
	\\
	To better derive information on the SZE-21cm produced by low redshift galaxy clusters, multi-frequency measurements of the SZE at X-ray and microwave bands are highly desirable for the extraction of additional information from its spectral shape, and for separating the signal from the various sources of contamination and confusion. Such multi-frequency measurements will also be useful in finding suitable cluster candidates for low-frequency studies based on the strength of the SZE signal at high-frequencies and will be instrumental in estimating the profile of the optical depth and the temperature of the cluster gas, which must be known when computing the SZE-21cm.

	\section{The SZE-21cm in the EDGES era}
	\label{sec:Edges}
	
		\begin{figure*}[t]
		\includegraphics[width=0.80\textwidth]{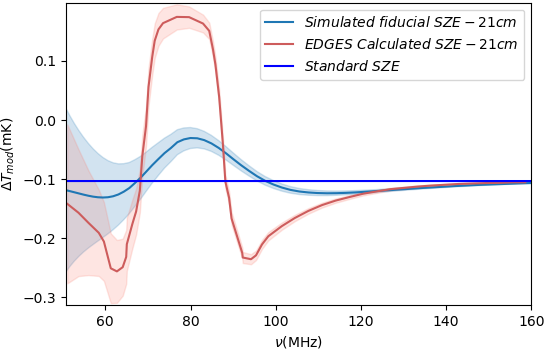}
		\centering
		\caption{The SZE-21cm computed for the simulated fiducial 21 cm background model from Figure \ref{fig:Tb} ($\textbf{\emph{Blue}}$) compared to the SZE-21cm computed using the 21cm background spectrum detection claimed by the EDGES experiment ($\textbf{\emph{Red}}$). The shaded errors regions assume observations with an interferometer similar to SKA for differential observations towards a single cluster. The SZE-21cm spectrum computed from the EDGES input spectrum is only reliable between 70-90 MHz due to the edge effects of the frequency range of the input spectrum. For both cases, we assume a cluster with a plasma temperature of 10 keV and an optical depth of $\tau = 10^{-3}$.}
		\label{fig:SZEcomp}
	\end{figure*}
	
\begin{figure*}[t]
		\includegraphics[width=1\textwidth]{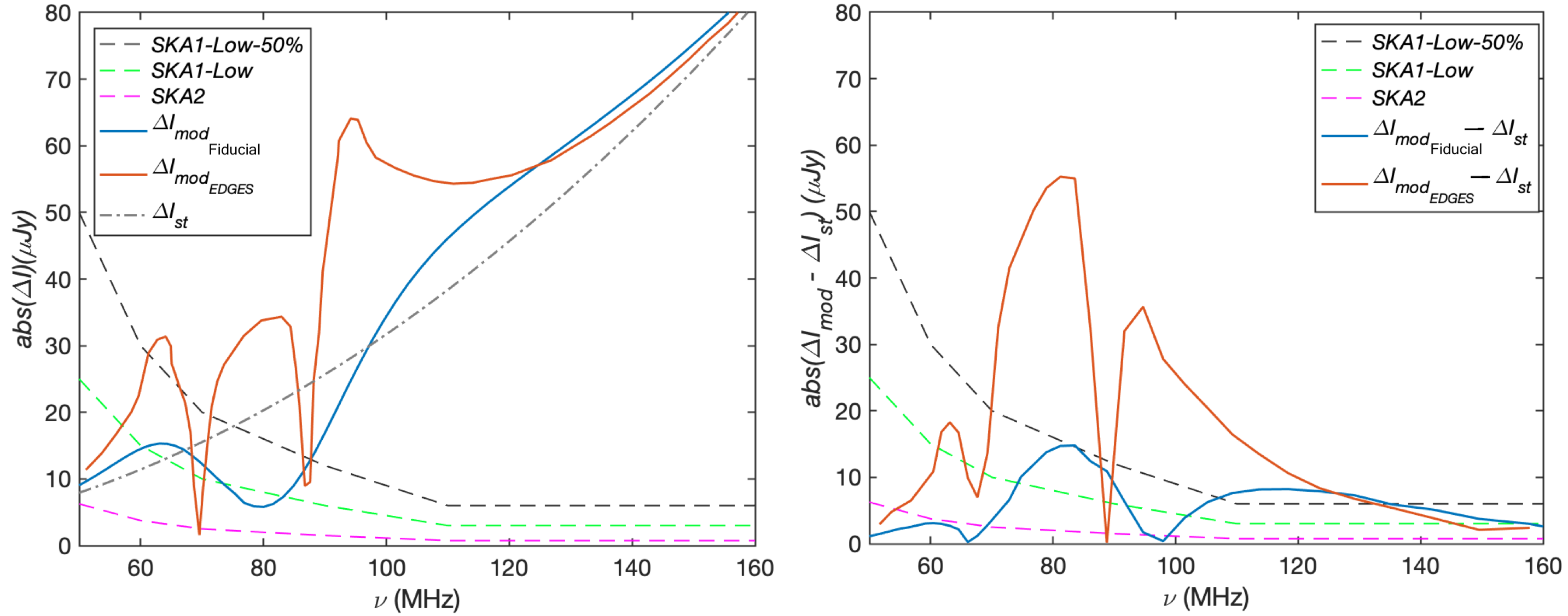}
		\centering
		\caption{\textbf{Right:} The spectra of the SZE-21cm $\Delta I_{mod}$. The \textbf{blue} solid line represents the SZE-21cm for our simulated fiducial 21 cm background model and the $\textbf{\emph{Red}}$ line for a 21 cm background spectrum similar to the EDGES detection. The $\textbf{\emph{Grey}}$ dotted and dashed line represents the SZE for a non-modified CMB $\Delta I_{st}$. \textbf{Left:} The absolute value of the difference between the SZE-21cm and the standard SZE. The $\textbf{\emph{Blue}}$ solid line for our simulated fiducial model and the $\textbf{\emph{Red}}$ line for the SZE-21cm calculated from the spectrum similar to the EDGES detection. For both figures, we compared with the SKA1-low-50$\%$ (\textbf{\emph{Black}}), SKA1-low (\textbf{\emph{Green}}), and SKA2 ($\textbf{\emph{Magenta}}$) sensitivities for 100 kHz bandwidth, 1000 hrs of integration, 2 polarizations, no taper, and no weight.}
		\label{fig:sense}
	\end{figure*}

	In this section, we compare our results for the SZE-21cm obtained using the fiducial 21 cm background model in section \ref{sec:standard} to the results we calculate using the 21 cm background spectrum claimed by the EDGES experiment \citep{Bowman2018}. To calculate the SZE-21cm with the input spectrum being the 21 cm background signal claimed by the EDGES collaboration we used the formalism in section \ref{subsec:sze21cm}. We show the spectra in Figure \ref{fig:SZEcomp}. We note that the SZE-21cm spectrum computed from the EDGES input spectrum is accurate only in the frequency range (70-90 MHz) due to the edge effects of the frequency range.
	\\
	\noindent
	The EDGES 21 cm background spectrum in Figure \ref{fig:Tb} implies that gas temperatures during DA and EoR were cooler by orders of 2 to 3 as compared to predictions by any prior models. The peak of the SZE-21cm for the EDGES detected background signal and the troughs of the SZE-21cm are also amplified by orders 2 to 3 compared to the theoretical model. Figure \ref{fig:SZEcomp} therefore, demonstrates that the SZE-21cm spectrum can be used to establish the amplitude of the input background spectrum and features that probe the corresponding epochs.
	\\
	In Figure \ref{fig:sense} we plot the low-frequency SKA telescope sensitivity compared to the standard SZE and the SZE-21cm. We note that the EGDES background is reliable only in the band 70-90 MHz. We list our findings below:
	
	\begin{itemize}
		\item With SKA1-low-50$\%$ the SZE-21cm can be detected for the fiducial model at $\nu$ $\gtrsim$ 90 MHz and the EDGE-like background at $\nu$ $\gtrsim$ 75 MHz. Good frequency channels to distinguish between the standard SZE and the SZE-21cm signals are 110 - 140 MHz for the fiducial model and 90 - 130 MHz for an EDGES-like background spectrum.
		
		\item With SKA1-low the SZE-21cm can be detected for the fiducial model at $\nu$ $\gtrsim$ 70 MHz and the EDGES-like background at $\nu$ $\gtrsim$ 60 MHz. Good frequency channels to distinguish between the standard SZE and the SZE-21cm signals is $\nu$ $\gtrsim$ 75 MHz for the fiducial model and $\nu$ $\gtrsim$ 70 MHz for an EDGES-like background spectrum.
		
		\item With SKA2 the SZE-21cm can be detected for both the fiducial model and EDGES-like background at $\nu$ $\gtrsim$ 50 MHz. Good frequency channels to distinguish between the standard SZE and the SZE-21cm signals is $\nu$ $\gtrsim$ 70 MHz for the fiducial model and $\nu$ $\gtrsim$ 55 MHz for an EDGES-like background spectrum.
	\end{itemize}
	
	\noindent
	To match the unexpectedly large extent of the SZE-21cm computed using the EDGES signal without recourse to any form of new physics we may need to change the conditions for the source of the background radiation in our simulated models. Apart from concerns raised on the foreground model and systematics, the EDGES detection has been discussed extensively and potential explanations consider new physics, radio background in excess of the CMB at redshifts related to the DA and EoR, interaction between baryons and dark-matter particles \citep{Barkana2018}, and new modes of star formation such as metal-free Pop III stars \cite[see e.g.,][]{Mebane2019}. While this detection has yet to be confirmed by other techniques and experiments, it exhibits features that are inconsistent with existing theoretical predictions. Independent experiments and methods are needed to verify this EDGES result, therefore, we propose measuring the SZE-21cm with upcoming instruments like the SKA as this method probes the features of the input 21 cm background spectrum.

	\section{Challenges for observations of the SZE-21cm}
	\label{sec:chal}
	
	\begin{figure*}[t]
		\includegraphics[width=0.6\textwidth]{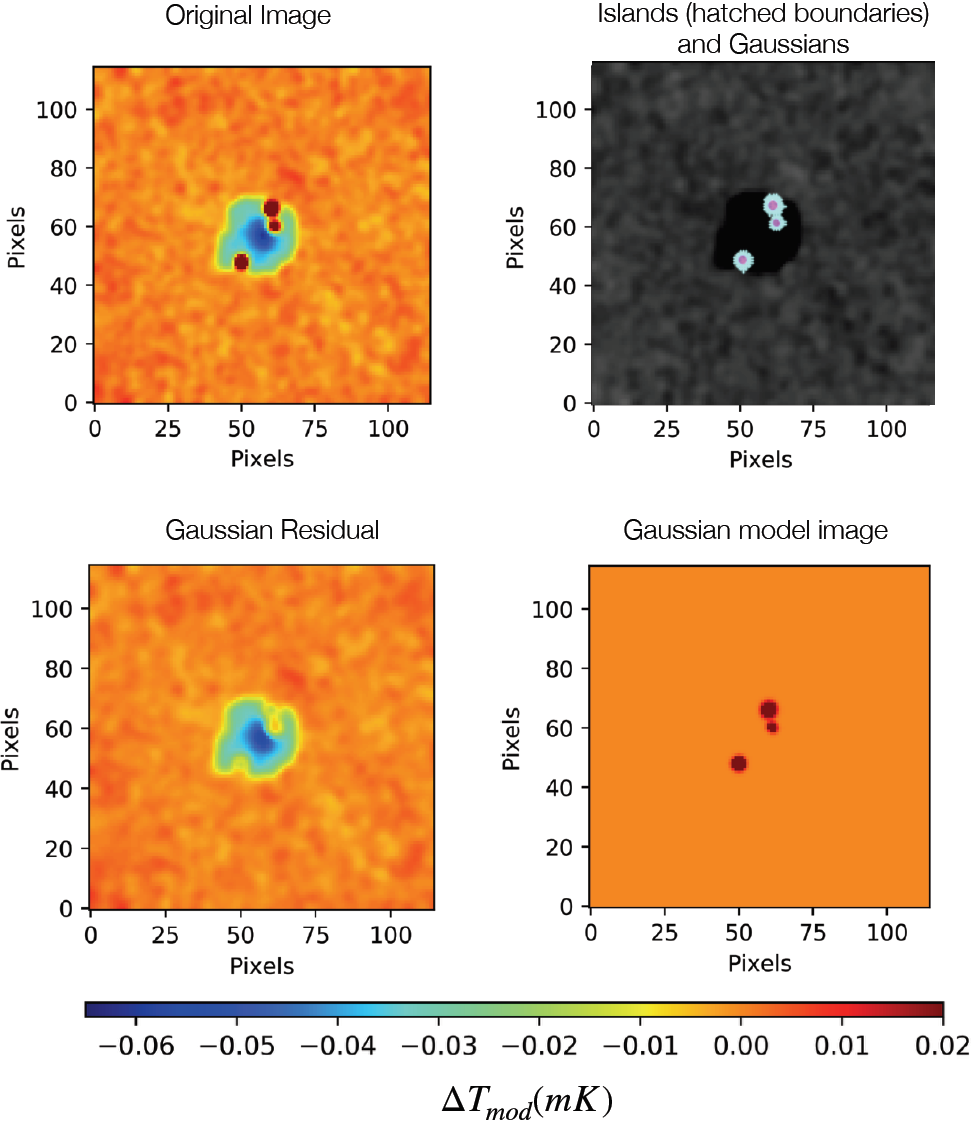}
		\centering
		\caption{Simulated SZE-21cm cluster maps at 176.67 MHz: with point sources in the original image,  the model image, and the residual which is original minus model image. Boundaries of the regions/island of emission found by code are shown in light blue and the fitted Gaussians the shaded areas shown for each region as small ellipses.}
		\label{fig:point}
	\end{figure*}
	
	Work by \cite{Cooray2006} and \cite{Colafrancesco2016} showed that contamination from background and foreground emissions on scales larger than a typical galaxy cluster has less impact on differential observational measures of the SZE-21cm. However, radio point sources have historically been the main source of contamination in the SZE measurements \citep{Carlstrom2002}, and synchrotron radio emissions from the host galaxy cluster can also lead to contamination at frequencies relevant to the DA and EoR studies. The measurements of the SZE-21cm can be contaminated by the emission of point sources along the sightline to the galaxy cluster, which could contribute to the underestimation or overestimation of the SZE signal. A strategy that could be used to remove point sources is to identify possible point sources in radio catalogs and follow up with pointed observations at high angular resolution. The point source removed should be much smaller than the extent of the SZE signal for the amount of the removed SZE-21cm flux to be negligible; this will be achievable with the SKA telescope, which should be capable of making high angular resolution measurements ($<$ 0.1 arcsecond). 
	\\
	\noindent
	Radio point sources are abundant in galaxy clusters, which leads to a significant fraction of the SZE signal being affected by the contamination as discussed above. Two modes can lead to the dilution of the SZE: point sources that are not removed can fill the SZE signal, and the removal of the point sources can result in removal of the SZE signal \citep{Cooray1998,Birkinshaw1999}. Experiments to measure the SZE only needs to remove about two bright point sources to be confident that the measurement is not contaminated by radio emission from galaxy cluster members \citep{Holder2002}. We simulated an observation towards a galaxy cluster at $z$ = 0.1 with point sources assuming the typical spectral index is about $\alpha = -0.7$, where $S(Jy) \propto \nu^{\alpha}$  at 176.67 MHz with radio flux ranging from 30 to 290 mJy \cite[see e.g.,][]{George2017}, which is of the order of magnitude of some of brightest galaxies in clusters at frequencies relevant to this work. We make use of a code designed to decompose radio interferometry images into sources called PyBDSF (the Python Blob Detector and Source Finder, formerly PyBDSM) \citep{Mohan2015}. The code reads in the input image, calculates background rms and mean images, finds regions of emission, fits Gaussians to the regions, and groups them into sources. We left the PyBDSF parameters at their default values during this source extraction.  We obtain a percentage error of 6\% for the SZE-21cm simulated for observation with the SKA1-low beam after point source subtraction compared to the SZE-21cm computed at the same frequency in Section \ref{sec:standard}. The presence of the point source and the subtraction thus has an impact on the SZE-21cm signal that we measure. The point source subtraction removes flux within the point source subtraction beam area from the cluster map at the position of the point source. Figure \ref{fig:point} illustrates the simulated point source and subtraction described here.
	\\
	\noindent
	Another possible issue is due to the presence of diffuse emission in the radio band (Radio halos, mini-halo and relics) that have been detected and studied in galaxy clusters \citep[see review,][]{Feretti2012}. This may be one order of magnitude higher at frequencies relevant to observations of the SZE-21cm. It is also worth noting that contamination due to synchrotron emission in the cluster decreases for objects at large distances, since radio synchrotron emission changes with luminosity distance, whereas the SZE does not change as a function of redshift and distance of the source \citep[see.,][]{Colafrancesco2016}. By using this property, Multi-frequency and spatially resolved analysis with upcoming interferometers will allow us to reduce the importance of intrinsic radio emission from the clusters. It is important, however, that we study and analyse both contributions to distinguish and separate them.
	
	\section{Conclusions}
	\label{sec:Conclusion}
	Further simulations and observations are necessary to pinpoint the best strategy for observing and measuring the SZE-21cm to probe the DA and EoR, which will require very deep, thorough observations and highly accurate theoretical analysis. The detection of the SZE-21cm with the upcoming HERA and SKA instruments has the potential to yield insightful information on the history and physical mechanisms occurring during the DA and the subsequent EoR as they are sensitive. The SKA will particularly play a large role in the study of cosmological radio backgrounds by providing SZE data with high accuracy. Cosmic variance and instrumental noise have potential to contaminate and decrease chances for a clean detection of the SZE-21cm signal. Averaging the signal from a number of clusters has been demonstrated as a method to suppress the noise and deal with the challenges we may face observing towards a individual cluster. To identify ideal clusters and obtain cluster plasma temperatures, X-ray measurements of the ICM properties will be of great importance and will form a crucial part of our observational strategy for studies of the SZE-21cm, particularly clusters that are in the Southern hemisphere, where HERA and the SKA will observe. In this regard, the next step would be to identify clusters in Microwave source catalogs from the SPT, Planck and, ACT. Furthermore, we have shown that point source removal is crucial to address the issue of contamination that may fill the SZE-21cm signal. We also noted a deviation between the true signal and simulated observation with HERA and SKA at high frequencies, which is particularly contrary to expectations for observations with SKA-low. The source of this deviation needs to be investigated and adequately addressed.  This work has shown that HERA and Low-frequency SKA will be able to obtain maps of the SZE-21cm. The outcomes from our simulations and calculations make us optimistic that the SKA telescope will detect the SZE-21cm spectrum and will be instrumental in following-up on the EDGES result.

		\acknowledgments
	The South African Radio Astronomy Observatory (SARAO), which is a facility of the National Research Foundation (NRF), a Department of Science and Innovation (DSI) agency, has sponsored this research. C.M., P.M. \& G. B. acknowledge the encouragement, advice and supervision of Prof. Sergio Colafrancesco in this work, may his soul rest in peace.


	%

\begin{thebibliography}{}
		\bibitem[\protect\citeauthoryear{Barkana}{2018}]{Barkana2018}
		Barkana R., 2018, Nature, 555, 71-74
		
		\bibitem[\protect\citeauthoryear{Birkinshaw}{1999}]{Birkinshaw1999}
		Birkinshaw M. 1999, Physics Reports, 310, 97
		
		\bibitem[\protect\citeauthoryear{Barkana \& Leob}{2005}]{Barkana2005}
		Barkana R.,  Leob A., 2005, ApJ, 626, 1
		
		\bibitem[\protect\citeauthoryear{Bowman et al.}{2009}]{Bowman2009}
		Bowman J. D., Morales M. F., Hewitt J. N., 2009, ApJ, 695, 183
		
		\bibitem[\protect\citeauthoryear{Bowman et al.}{2018}]{Bowman2018}
		Bowman J. D., Rogers A. E. E., Monsalve R. A., Mozdzen T. J.,  Mahesh N., 2018, Nature (London), 555, 67
		
		\bibitem[\protect\citeauthoryear{Bradley et al.}{2019}]{Bradley2019}
		Bradley R. F., Tauscher K., Rapetti D., Burns J.O., 2019, The Astrophysical Journal, 874, 153
		
		\bibitem[\protect\citeauthoryear{Bromm et al.}{2009}]{Bromm2009}
		Bromm V., Yoshida N., Hernquist L., McKee C. F., 2009, Nature, 459, 49-54
		
		\bibitem[\protect\citeauthoryear{Burns et al.}{2011}]{Burns2011}
		Burns J. O., Lazio J., Bowman J., Bradley R., et al., 2011,  Advances in Space Research, 218, 233.03
		
		\bibitem[\protect\citeauthoryear{Carlstrom et al.}{2002}]{Carlstrom2002}
		Carlstrom J.E.,  Holder G.P., Reese E.D., 2002, Ann. Rev. Astron. Astrophys, 40, 643-680 (2002)
		
		\bibitem[\protect\citeauthoryear{Cavaliere \& Fusco-Femiano}{1976}]{Cavaliere1976}
		Cavaliere A., \& Fusco-Femiano R. 1976, A \& A, 49, 137
		
		\bibitem[\protect\citeauthoryear{Clanton et al.}{2012}]{Clanton2012}
		Clanton C., Beichman C., Vasisht G. Smith R. Gaudi B. S., 2012, The Astronomical Society of the Pacific, vol. 124, 917
		
		
		\bibitem[\protect\citeauthoryear{Colafrancesco et al.}{2002}]{Colafrancesco2002}
		Colafrancesco S., Marchegiani, P., Palladino, E. 2002, A\&A, 397, 27
		
		\bibitem[\protect\citeauthoryear{Colafrancesco et al.}{2016}]{Colafrancesco2016}
		Colafrancesco S.,  Marchegiani P., Emritte M. S., 2016, A\&A, 595, A21 (2016)
		
		
		\bibitem[\protect\citeauthoryear{Cooray}{1998}]{Cooray1998}
		Cooray A. R., Grego L., Holzapfel W. L., Joy M., Carlstrom J. E., 1998, AJ, 115, 1388
		
		\bibitem[\protect\citeauthoryear{Cooray}{2006}]{Cooray2006}
		Cooray A. R., 2006, Phys. Rev. D, 73, 103001
		
		\bibitem[\protect\citeauthoryear{DeBoer}{2017}]{DeBoer2017}
		DeBoer D. R., Parsons A. R., Aguirre J. E., et al., 2017, Publications of the Astronomical Society of the Pacific, 129, 974, 045001
		
		\bibitem[\protect\citeauthoryear{En{\ss}lin \& Kaiser}{2000}]{Ensslin2000}
		En{\ss}lin T. A., Kaiser C. R., 2006, A \& A, 360, 417-430
		
		\bibitem[\protect\citeauthoryear{Ewall-Wice et al.}{2018}]{Ewall-Wice2018}
		Ewall-Wice A., Chang T.C., Lazio  J., Doré O., Seiffert M., and Monsalve R.A., 2018, ApJ, 868, 63
		
		\bibitem[\protect\citeauthoryear{Ewen \& Purcell}{1951}]{Ewen1951}
		Ewen H., Purcell E., 1951, Nature, 168, 356
		
		\bibitem[\protect\citeauthoryear{Fan}{2006}]{Fan2006}
		Fan X., 2006, AJ, 132, 117
		
		\bibitem[\protect\citeauthoryear{Feretti et al.}{2012}]{Feretti2012}
		Feretti L., Giovannini G., Govoni F., et al., 2012, Astron Astrophys Rev, 120, 54
		
		\bibitem[\protect\citeauthoryear{Fialkov et al.}{2018}]{Fialkov2018}
		Fialkov A., Barkana, R., and Cohen, A., 2018, PhRvL, 121, 011101
		
		\bibitem[\protect\citeauthoryear{Field}{1958}]{Field1958}
		Field G., 1958, ApJ, 129, 536
		
		\bibitem[\protect\citeauthoryear{Fixsen}{2011}]{Fixsen2011}
		Fixsen D.J., Kogut A., Levin S., Limon M., Lubin P., Mirel P., Seiffert M., Singal J., Wollack E., Villela T., Wuensche C.A., 2011, Astrophys. J., 734, p. 5
		
		\bibitem[\protect\citeauthoryear{Furlanetto}{2006}]{Furlanetto2006}
		Furlanetto S. R., 2006, MNRAS, 371, 867-878
		
		
		\bibitem[\protect\citeauthoryear{George et al.}{2017}]{George2017}
		George L.  T., Dwarakanath K. S., Johnston-Hollitt M., 2017, MNRAS, 467, 936-949
		
		\bibitem[\protect\citeauthoryear{Hassan et al.}{2018}]{Hassan2018}
		Hassan S., Liu A., Kohn S., La Plante P., 2018, MNRAS, 483, 2524
		
		\bibitem[\protect\citeauthoryear{Hills et al.}{2018}]{Hills2018}
		Hills R., Kulkarni G., Meerburg P.D., et al., 2018, Nature, 564, E32–E34
		
		\bibitem[\protect\citeauthoryear{Hansen et al.}{2002}]{Hansen2002}
		Hansen S. H., Pastor S., Semikoz V. S., 2002, ApJ, 573, L69–L71
		
		\bibitem[\protect\citeauthoryear{Holder et al.}{1999}]{Holder1999}
		Holder G. P.,Carlstrom J. E., 1999, ASP Conf. Ser., 181, 199
		
		\bibitem[\protect\citeauthoryear{Holder}{2002}]{Holder2002}
		Holder G. P., 2002, The Astrophysical Journal, 580,36–41
		
		\bibitem[\protect\citeauthoryear{Jacobs et al.}{2015}]{Jacobs2015}
		Jacobs D. C., 2015, The Astrophysical Journal, 801, 1, 51
		
		\bibitem[\protect\citeauthoryear{Jacobs et al.}{2016}]{Jacobs2016}
		Jacobs D. C., 2016, The Astrophysical Journal, 825, 2, 114
		
		
		\bibitem[\protect\citeauthoryear{Liu et al.}{2009}]{Liu2009}
		Liu A., Tegmark M., Zaldarriaga M. 2009, MNRAS, 394, 1575
		
		\bibitem[\protect\citeauthoryear{Liu et al.}{2014}]{Liu2014}
		Liu A., Parsons A. R.,Trott C. M., et al., 2014, Phys. Rev. D., 90, 023018
		
		\bibitem[\protect\citeauthoryear{Liu et al.}{2016}]{Liu2016}
		Liu A., Pritchard J. R., Allison R., et al., 2016, Phys. Rev. D., 93, 043013
		
		\bibitem[\protect\citeauthoryear{Mebane et al.}{2019}]{Mebane2019}
		Mebane R.H., Mirocha J., Furlanetto S.R., 2020, MNRAS, 493, 1, 1217-1226
		
		\bibitem[\protect\citeauthoryear{Mellema et al.}{2012}]{Mellema2012}
		Mellema G., Koopmans L., Abdalla F., Bernardi G., Ciardi B., Daiboo S., de Bruyn G., Datta K.K., Falcke H., Ferrara A., et al., 2012, ArXiv e-prints:1210.0197
		
		\bibitem[\protect\citeauthoryear{McKinley et al.}{2018}]{McKinley2018}
		McKinley B., Bernardi G., Trott C.M., Line J.L.B., Wayth R.B., Offringa A.R., Pindor B., Jordan C.H., Sokolowski M., Tingay S.J., et al., 2018, MNRAS, 481, 5043-5045
		
		\bibitem[\protect\citeauthoryear{McQuinn et al.}{2006}]{McQuinn2006}
		McQuinn M., et al., 2006, ApJ, 653, 815
		
		
		\bibitem[\protect\citeauthoryear{Mesinger et al.}{2010}]{Mesinger2010}
		Mesinger A., Furlanetto S.R., Cen, R., 2010, MNRAS, 411, 955-972
		
		\bibitem[\protect\citeauthoryear{Mondal et al.}{2019}]{Mondal2019}
		Mondal R., Shaw A. K., Iliev I. T., Bharadwaj S., Datta K. K., 2019, MNRAS, 494, 3, 4043-4056
		
		\bibitem[\protect\citeauthoryear{Mohan \& Rafferty}{2015}]{Mohan2015}
		Mohan N., Rafferty D., 2015, PyBDSM: Python Blob Detection and Source Measurement, Astrophysics Source Code Library
		
		\bibitem[\protect\citeauthoryear{Muller \& Oort}{1957}]{Muller1957}
		Muller C., Oort J., 1957, Bulletin of the Astronomical Institutes of the Netherlands, 13, 151
		
		\bibitem[\protect\citeauthoryear{Paciga et al.}{2013}]{Paciga2013}
		Paciga G., Albert J. G., Bandura K., et al. 2013, MNRAS, 443, 639-647
		
		\bibitem[\protect\citeauthoryear{Planck Collaboration}{2015}]{Planck2015}
		Planck Collaboration Ade P. A. R., Aghanim N., Arnaud M., Ashdown M., Aumont J., Baccigalupi C., Banday A. J., Barreiro R. B., et al., 2015, ArXiv e-prints:1502.01589
		
		\bibitem[\protect\citeauthoryear{Pober et al.}{2013}]{Pober2013}
		Pober J. C., Parsons A. R., DeBoer D. R., et al. 2013, AJ, 145, 65
		
		\bibitem[\protect\citeauthoryear{Pober et al.}{2014}]{Pober2014}
		Pober J. C., Liu A., Dillon J. S., et al. 2014, ApJ, 782, 66
		
		\bibitem[\protect\citeauthoryear{Pritchard \& Leob}{2012}]{Pritchard2012}
		Pritchard J. R., and Loeb A., 2012, Rep. Prog. Phys. 75, 086901
		
		\bibitem[\protect\citeauthoryear{Rephaeli}{1995}]{Rephaeli1995}
		Rephaeli Y., 1995, Annual Review of Astronomy and Astrophysics, 33, 1, 541-579
		
		\bibitem[\protect\citeauthoryear{Romer et al.}{2001}]{Romer2001}
		Romer A. K., et al, 2001, The Astrophysical Journal, 547, 2, 594-608
		
		\bibitem[\protect\citeauthoryear{Santos et al.}{2005}]{Santos2005}
		Santos, M., Cooray, A., and Knox, L., 2005, ApJ, 625, 575-587
		
		\bibitem[\protect\citeauthoryear{Shaver et al}{1999}]{Shaver1999}
		Shaver P.A., Windhorst R.A., Madau P., and de Bruyn A.G., 1999, A\&A, 345, 380–390
		
		\bibitem[\protect\citeauthoryear{Sims \& Pober}{2019}]{Sims2019}
		Sims P.H., and Pober J.C., 2019, MNRAS, 488, 2904–2916
		
		\bibitem[\protect\citeauthoryear{Sims \& Pober}{2019}]{Sims22019}
		Sims P.H., and Pober J.C., 2019, MNRAS, 492, 22–38
		
		\bibitem[\protect\citeauthoryear{Singh \& Subrahmanyan}{2019}]{Singh2019}
		Singh S., Subrahmanyan R., 2019, The Astrophysical Journal, 880, 26
		
		\bibitem[\protect\citeauthoryear{Sunyaev \& Zel'dovich}{1970}]{SZ1}
		Sunyaev M., Zel'dovich Y., 1970, Comments Astrophys. Space Phys., 2, 66
		
		\bibitem[\protect\citeauthoryear{Sunyaev \& Zel'dovich}{1972}]{SZ2}
		Sunyaev M., Zel'dovich Y., 1972, Comments Astrophys. Space Phys., 4, 173
		
		\bibitem[\protect\citeauthoryear{Sunyaev \& Zel'dovich}{1980}]{SZ}
		Sunyaev M., Zel'dovich Y., 1980, Ann. Rev. Astron. Astrophys, 18, 537
		
		\bibitem[\protect\citeauthoryear{Takalana et al.}{2018}]{Takalana2018}
		Takalana C. M., Colafrancesco S.,  Marchegiani P., 2018, PoS HEASA2017 (2018) 008
		
		\bibitem[\protect\citeauthoryear{van de Hulst}{1945}]{vandeHulst1945}
		van de Hulst H. C., 1945, Ned.Tijd.Natuurkunde, vol.11, 210
		
		\bibitem[\protect\citeauthoryear{Wise}{2019}]{Wise2019}
		Wise J. H., 2019, Contemp. Phys., 60, 145
		
		
		\bibitem[\protect\citeauthoryear{Zaldarriaga}{2004}]{Zaldarriaga2004}
		Zaldarriaga M., Furlanetto S., Hernquist L., 2004, ApJ, 608, 622
		
		\bibitem[\protect\citeauthoryear{Zaroubi}{2013}]{Zaroubi2013}
		Zaroubi S., 2013, The First Galaxies - Theoretical Predictions and Observational Clues, Springer, vol 396
		
		\bibitem[\protect\citeauthoryear{Zel'dovich \& Sunyaev}{1969}]{SZ3}
		Zel'dovich Y., Sunyaev, R., 1969, Astrophys. Space Sci., 4, 301
		
		
	\end{thebibliography}

	%

\end{document}